\documentclass{article}
\newenvironment{keywords}{\begin{quote}{\bf Keywords:}}{\end{quote}}
\newcommand\PARstart[2]{#1#2}
\newenvironment{proof}{{\sl Proof:}}{\hfill{$\square$}}
\newcommand\IEEEmembership[1]{\ignorespaces}

\usepackage{amssymb}
\usepackage{amsmath}
\usepackage{graphics}

\newtheorem{theorem}{Theorem}
\newtheorem{lemma}[theorem]{Lemma}

\renewcommand{\leq}{\leqslant}
\renewcommand{\geq}{\geqslant}
\def\N{{\mathbb N}}

\def\F{{\mathbb F}}

\def\mF{{\mathcal F}}

\def\phi{\varphi}

\def\Lam{\Lambda}

\def\lam{\lambda}

\def\del{\delta}

\newcommand\mut[1]{\ignorespaces}

\begin{document}

\title{The Order Bound on the Minimum Distance of the One-Point Codes
Associated to a Garcia-Stichtenoth Tower of Function Fields}

\author{Maria Bras-Amor\'os, Michael E. O'Sullivan,~\IEEEmembership{Member,~IEEE,}
\thanks{M. Bras-Amor\'os is with Universitat Aut\`onoma de Barcelona
  (e-mail: mbras@deic.uab.cat)}
\thanks{M. E. O'Sullivan is with San Diego State University (e-mail: mosulliv@sciences.sdsu.edu)}
}

\markboth{IEEE Transactions on Information Theory, VOL. X,
  NO. X}{BRAS-AMOR\'OS AND O'SULLIVAN: THE ORDER BOUND ON
  THE MINIMUM DISTANCE OF ONE-POINT CODES ASSOCIATED TO A GARCIA-STICHTENOTH TOWER OF FUNCTION FIELDS}

\maketitle

\begin{abstract}
Garcia and Stichtenoth discovered two towers of function fields that
meet the Drinfeld-Vl\u adu\c t bound on the ratio of  the number of
points to the genus. For one of these towers, Garcia, Pellikaan and
Torres derived a recursive description of the  Weierstrass
semigroups associated to a tower of points on the associated curves.
In this article, a non-recursive description of the semigroups is
given and from this the enumeration of each of the semigroups is
derived as well as its inverse. This enables us to find an explicit formula for the order
(Feng-Rao) bound on the minimum distance of the associated one-point
codes.
\end{abstract}


\section*{Introduction}

\PARstart{L}{et} $\N_0$ denote the set of all non-negative integers. A {\it
numerical semigroup} is a subset $\Lambda$ of $\N_0$ containing $0$,
closed under summation and with finite complement in $\N_0$.
The {\it enumeration} of $\Lambda$ is the unique increasing bijective map
$\lambda:\N_0\longrightarrow\Lambda$.
Usually $\lambda_i$ is used instead of $\lambda(i)$.
Given a numerical semigroup $\Lambda$ with enumeration $\lambda$
define the sequence $\nu_i$ by
$$\nu_i=\lvert\{j\in\N_0:\lambda_i-\lambda_j\in\Lambda\}\rvert.$$
This sequence
has become of great importance in
the theory of one-point algebraic-geometry codes.
For a sequence of one-point codes on a curve, where the one point is $P$,
the numerical semigroup considered is that
of the pole orders in $P$ of functions having only poles in $P$, usually named as the {\it Weierstrass semigroup} of $P$.
The sequence $\nu_i$ is used to define the {\it order bound} on the
minimum distance of one-point algebraic-geometry codes:
$$\delta_i=\min\{\nu_j:j>i\}.$$
The order bound, also known as Feng-Rao bound, is a lower bound on
the minimum distance of the $i$-th one-point code on $P$. In this
case the numerical semigroup is the Weierstrass semigroup associated
to $P$. Details can be found in
\cite{FeRa:dFR,HoLiPe:agc,KiPe:telescopic}.

Garcia and Stichtenoth first gave in
 \cite{GaSt:attainingDV,GaSt:tff}
 two towers of function fields
 attaining the
 Drinfeld-Vl\u adu\c t bound,
 which became of great importance in the area of algebraic coding theory.
 The tower in \cite{GaSt:tff}
  is defined over the finite field with $q^2$ elements $\F_{q^2}$
 for $q$ a prime power. It is given by
 \begin{itemize}
 \item $\mF^1=\F_{q^2}(x_1)$
 \item $\mF^m=\mF^{m-1}(x_m)$
 with $x_m$
 satisfying
 $$x_m^q+x_m=\frac{x_{m-1}^{q}}{x_{m-1}^{q-1}+1}.$$
 \end{itemize}
 It is shown in
 \cite{GaSt:tff}
 that the number of its rational points is
 $N_q(\mF^m)\geq(q^2-q)q^{m-1}$
 and that the genus $g_m$ of $\mF^m$
 is
 $g_m=
 (q^{\lfloor\frac{m+1}{2}\rfloor}-1)
 (q^{\lceil\frac{m-1}{2}\rceil}-1).
 $
 Hence, the ratio between
 the
 genus
 $g(\mF^m)$
 and
 $N_{q^2}(\mF^m)$
 converges to
 $1/(q-1)$, the Drinfeld-Vl\u adu\c t bound, as $m$ increases.
From these curves one can construct asymptotically good sequences of
codes.

For every function field $\mF^m$
in the tower
we distinguish
the rational point $Q^m$
that is the unique pole of $x_1$.
\mut{Then we can
consider the one-point codes
defined on this distinguished point,
which
we call
{\it codes on the tower of Garcia-Stichtenoth}.
We know that for
algebraic-geometry codes
the dimension $k$, the minimum distance $d$
and the length $n$ satisfy
$k+d\geq n+1-g$
where $g$ is the genus of the defining function field.
So, dividing by $n$,
we get a relationship
between the information rate $R$
and the relative minimum distance $\delta$:
$R+\delta\geq 1-\frac{g-1}{n}.$
Now, using this inequality
we can see that
for a sequence of
function fields attaining the
Drinfeld-Vl\u adu\c t
bound,
we can find a
sequence of
algebraic-geometry codes
with
limit value of the information rate equal to a
non-zero
constant
$R$,
and
limit value of the relative minimum distance equal to a
non-zero constant
$\delta$,
such that
$R+\delta\geq 1-\frac{1}{q-1}.$
Notice that this is the
Tsfasman-Vl\u adu\c t-Zink bound.
}
The Weierstrass semigroup $\Lambda^m$ at
$Q^m$ in $\mF^m$
was recursively described
in \cite{PeStTo}.
Indeed,
the semigroups are given
recursively by
\begin{equation}
\label{eq:semigroups}
\begin{array}{rcl}
\Lambda^1&=&\N_0
\\
\Lambda^{m}&=&q\cdot\Lambda^{m-1}\cup\{i\in\N_0: i\geq
q^m-q^{\lfloor\frac{m+1}{2}\rfloor}\}.
\end{array}
\end{equation}

Although these Weierstrass semigroups have been known for a long time,
no explicit description of the order bound on the minimum distance
for the associated one-point codes
has appeared in the literature.
Chen found certain bounds on the order bound for a given range
and this enabled him to prove that some codes had distance
larger than the order bound \cite{Chen}.
The main goal of this paper is
to find the explicit description of the order bound
from a deep analysis of these semigroups.


In Section~\ref{sec:gs2-semigroup}
we give a non-recursive description of these semigroups.
This description leads to explicit formulation of
their enumerations as well as for the inverses of their enumerations.
This is presented in Section~\ref{sec:gs2-enumeration}.
In Section~\ref{sec:obound}
we find the explicit formula for the order bound on the minimum distance
of the one-point codes associated to the tower of function fields.

\section{Non-recursive description of the semigroups}
\label{sec:gs2-semigroup}

In this section we will give a non-recursive description of the
semigroups (\ref{eq:semigroups}).
The {\it conductor} of a numerical semigroup $\Lambda$
is the unique integer $c$ such that $c-1\not\in\Lambda$ and
$c+\N_0\subseteq\Lambda$.
The {\it gaps} of $\Lambda$ are the elements in
$\N_0\setminus\Lambda$
and the {\it non-gaps}
are the elements in $\Lambda$.

From now on we will use
$c_m$ for the conductor of $\Lambda^m$, which is $q^m-q^{\lfloor\frac{m+1}{2}\rfloor}$, and
 $A_i=\{c_{2i-1}+j: j=0,\dots,q^{i-1}(q-1)-1\}=\{j\in\N_0:q^{2i-1}-q^i\leq j\leq q^{2i-1}-q^{i-1}-1\}$.
Notice that these are not
recursive definitions.

\begin{theorem}
\label{theorem:gs2-semigroup}
$$\Lambda^m=\bigsqcup_{i=1}^{\lfloor\frac{m}{2}\rfloor}q^{m-2i+1}A_{i}\sqcup\{j\in\N_0:
j\geq c_m\}.$$
\end{theorem}

\begin{proof}
We will proceed by induction.
The case $m=1$ is obvious.
Suppose $m>1$.
If $m$ is even, say $m=2n$, then
$$\Lambda^{2n}=q\Lambda^{2n-1}\cup\{i\in\N_0: i\geq c_{2n}\}.$$
By the induction hypothesis, since $c_{2n-1}=q^{2n-1}-q^{n}$ and $c_{2n}=q^{2n}-q^{n}$,
\begin{align*}
\Lambda^{2n}&=\cup_{i=1}^{n-1}q^{2n-2i+1}A_i\cup q\{j\in\N_0: j\geq q^{2n-1}-q^{n}\}\\&\hfill{\cup\{j\in\N_0: j\geq q^{2n}-q^{n}\}}\\
&=\cup_{i=1}^{n-1}q^{2n-2i+1}A_i\\&\hfill{\cup q\{j\in\N_0: q^{2n-1}-q^{n}\leq j\leq q^{2n-1}-q^{n-1}-1\}}\\&\hfill{\cup\{j\in\N_0: j\geq q^{2n}-q^{n}\}}\\
&=\cup_{i=1}^{n}q^{2n-2i+1}A_i\cup\{j\in\N_0: j\geq q^{2n}-q^{n}\}\\
&=\cup_{i=1}^{\lfloor\frac{(2n)}{2}\rfloor}q^{(2n)-2i+1}A_i\cup\{j\in\N_0: j\geq c_{2n}\}.\\
\end{align*}

If $m$ is odd, say $m=2n+1$, then
$$\Lambda^{2n+1}=q\Lambda^{2n}\cup\{i\in\N_0: i\geq c_{2n+1}\}.$$
By the induction hypothesis, since $c_{2n}=q^{2n}-q^{n}$ and $c_{2n+1}=q^{2n+1}-q^{n+1}$,
\begin{align*}
\Lambda^{2n+1}&=\cup_{i=1}^{n}q^{2n-2i+2}A_i\cup q\{j\in\N_0: j\geq q^{2n}-q^{n}\}\\&\hfill{\cup\{j\in\N_0: j\geq q^{2n+1}-q^{n+1}\}}\\
&=\cup_{i=1}^{n}q^{2n-2i+2}A_i\cup\{j\in\N_0: j\geq q^{2n+1}-q^{n+1}\}\\
&=\cup_{i=1}^{\lfloor\frac{(2n+1)}{2}\rfloor}q^{(2n+1)-2i+1}A_i\cup\{j\in\N_0: j\geq c_{2n+1}\}.\\
\end{align*}

It remains to see that the unions are disjoint.
Let us first see that the sets $q^{m-2i+1}A_{i}$
are pairwise disjoint.
Indeed,
\begin{eqnarray*}
\mbox{max}\left(q^{m-2i+1}A_{i}\right)
&=&q^{m-2i+1}(q^{2i-1}-q^{i-1}-1)\\
&=&q^m-q^{m-i}-q^{m-2i+1}.
\end{eqnarray*}

On the other hand,
\begin{eqnarray*}
\mbox{min}\left(q^{m-2(i+1)+1}A_{i+1}\right)
&=&q^{m-2i-1}(q^{2i+1}-q^{i+1})\\
&=&q^m-q^{m-i}.
\end{eqnarray*}
This proves
that the sets $q^{m-(2i-1)}A_{i}$
are pairwise disjoint.

Now,
the maximum
attained by these sets is
$q^m-q^{m-\lfloor\frac{m}{2}\rfloor}-q^{m-2\lfloor\frac{m}{2}\rfloor+1}$.
If $m$ is even then this equals $q^m-q^{\frac{m}{2}}-q=c_m-q$
while if $m$ is odd then
this equals
$q^m-q^{\frac{m+1}{2}}-q^2=c_m-q^2.$
\end{proof}

\section{Enumeration of the semigroups}
\label{sec:gs2-enumeration}

In this section we use the notations and results
in Theorem~\ref{theorem:gs2-semigroup}
to find an explicit formula for the enumeration of the semigroups in (\ref{eq:semigroups})
as well as a formula for its inverse.

\begin{theorem}
\label{theorem:gs2-enumeration}
Let $\lambda$ be the enumeration
of $\Lambda^m$.
\begin{enumerate}
\item
The conductor $c_m$ is the image by $\lambda$ of $q^{\lfloor\frac{m}{2}\rfloor}-1$. That is,
$c_m=\lambda_{q^{\lfloor\frac{m}{2}\rfloor}-1}$.
\item
If $t\geq q^{\lfloor\frac{m}{2}\rfloor}-1$, then
$\lambda_t=c_m+t-q^{\lfloor\frac{m}{2}\rfloor}+1.$
\item
If $0\leq t< q^{\lfloor\frac{m}{2}\rfloor}-1$, then
$\lambda_t=q^{m-2l-1}
(c_{2l+1}+t+1-q^l),$
where $l=\lfloor\log_q(t+1)\rfloor$.
\end{enumerate}
\end{theorem}

\begin{proof}
\mbox{}
\begin{enumerate}
\item
For any positive integer $i$ it holds
$\lvert A_{i}\rvert=q^{i-1}(q-1)$ and hence
\begin{eqnarray*}
\lvert \bigsqcup_{i=1}^{\lfloor\frac{m}{2}\rfloor}q^{m-2i+1}A_{i}\rvert&=&
(1+q+q^2+\dots+q^{\lfloor\frac{m}{2}\rfloor-1})(q-1)\\&=&q^{\lfloor\frac{m}{2}\rfloor}-1.
\end{eqnarray*}
Since $c_m$ is the first non-gap which is not in
$\bigsqcup_{i=1}^{\lfloor\frac{m}{2}\rfloor}q^{m-2i+1}A_{i}$,
$c_m=\lambda_{q^{\lfloor\frac{m}{2}\rfloor}-1}$.

\item
If $t\geq q^{\lfloor\frac{m}{2}\rfloor}-1$, then $\lambda_t$ is in the subset
$\{j\in\N_0: j\geq c_m\}$ and for all $\lambda_k$ in this subset
we have,
$\lambda_{k+l}=\lambda_k+l\mbox{ for all }l\geq 0.$
Taking $k=q^{\lfloor\frac{m}{2}\rfloor}-1$ and
$l=t-q^{\lfloor\frac{m}{2}\rfloor}+1$
we get
$\lambda_t=\lambda_{q^{\lfloor\frac{m}{2}\rfloor}-1}+t-q^{\lfloor\frac{m}{2}\rfloor}+1=
c_m+t-q^{\lfloor\frac{m}{2}\rfloor}+1.$

\item
If $t< q^{\lfloor\frac{m}{2}\rfloor}-1$, then the lemma below shows
that $t$ may be uniquely expressed as $t= q^{l-1} + j-1$  where $l=
\lfloor \log_q (t+1) \rfloor +1\leq \lfloor m/2 \rfloor$ and $0 \leq
j \leq q^{l-1}(q-1)-1$. Since $\lvert
\bigsqcup_{k=1}^{l-1}q^{m-2k+1}A_{k}\rvert= q^{l-1}-1$, we have
$\lam_{q^{l-1} -1}$ is the first non-gap which is not in
$\bigsqcup_{k=1}^{l-1}q^{m-2k+1}A_{k}$, namely $q^{m-2l+1}c_{2l-1}$.
The next non-gaps after $q^{m-2l+1}c_{2l-1}$ are
$q^{m-2l+1}(c_{2l-1}+ i)$ with $i=1,\dots,q^{i-1}(q-1)-1$. Therefore,
$\lam_t = q^{m-2l+1}(c_{2l-1}+j) = q^{m-2l+1}(c_{2l-1}+t +1 - q^l).$
\end{enumerate}
\end{proof}

\begin{lemma}
\label{lem}
Any nonnegative integer $t$ may be uniquely expressed in the form
\[
t = q^{l-1} + j -1
\]
for $1 \leq l$ and $0 \leq j \leq q^{l-1} (q-1) -1$.
Furthermore, $l= \lfloor \log_q(t+1) \rfloor$ and $j = t - q^{l-1}
+1$.
\end{lemma}

\begin{proof}
Observe that
\begin{align*}
l = \lfloor \log_q (t+1)\rfloor +1  &\iff q^{l-1}  \leq t+1 < q^l  \\
 &\iff 0 \leq t -q^{l-1} +1 < q^{l-1} (q-1)  \\
\end{align*}
Setting $ j=  t -q^{l-1} +1$  shows that $t$ may be expressed as
claimed, and that the choice of $l$ and $j$ are unique.
\end{proof}

We want to find now a formula for the inverse of the enumeration of a numerical semigroup.
Given an integer $k$ and a numerical semigroup $\Lambda$
we define the {\it semigroup floor} of $k$ with respect to $\Lambda$
as the largest element in $\Lambda$ which is not larger than $k$.
It is denoted $\lfloor k\rfloor_\Lambda$.
In the next theorem we describe a formula not only to find the inverse
of the enumeration of a numerical semigroup, but also
to find the index for the semigroup floor of any integer.

\begin{theorem}
\label{theorem:inverseenumeration}
Let $\lambda$ be the enumeration of $\Lambda^m$ and let $k\geq 0$ be an integer.
Let $c_i$ and $g_i$ be respectively the conductor and the genus of $\Lambda^i$. Then,
$$
\lambda^{-1}(\lfloor k \rfloor_{\Lambda^m})=
\left\{\begin{array}{ll}
k-g_m
&\mbox{ if }k\geq c_m,
\\
q^{l-1}-1+\lfloor\frac{k}{q^{m-2l+1}}\rfloor-c_{2l-1}
&\mbox{ if }k< c_m,
\\
\end{array}\right.
$$
where
$l=m+1-\lceil\log_q(q^m-k)\rceil$.
\end{theorem}

\begin{proof}
The result for the case
when $k\geq c_m$ is obvious.
Suppose $k<c_m$.
By Theorem~\ref{theorem:gs2-semigroup},
$\lfloor k \rfloor_{\Lambda^m}$
can be expressed as
$q^{m-2l+1}(c_{2l-1}+ i)$ with $i=1,\dots,q^{i-1}(q-1)-1$
and with $l$ being the largest integer $l$
with
$q^{m-2l+1}c_{2l-1}\leq
\lfloor k \rfloor_{\Lambda^m}$.
Notice that it is also
the largest integer $l$
with
$q^{m-2l+1}c_{2l-1}\leq
k$.
By substituting $c_{2l-1}$
by $q^{2l-1}-q^l$
one gets
$l=m+1-\lceil\log_q(q^m-k)\rceil$.
By Theorem~\ref{theorem:gs2-enumeration}. 3), it follows that
$\lambda^{-1}(\lfloor k \rfloor_{\Lambda^m})=q^{l-1}-1+\frac{\lfloor k \rfloor_{\Lambda^m}}{q^{m-2l+1}}-c_{2l-1}$.
Now,
$\frac{\lfloor k \rfloor_{\Lambda^m}}{q^{m-2l+1}}=\lfloor\frac{k}{q^{m-2l+1}}\rfloor$
is a consequence of the fact that
$\lfloor k \rfloor_{\Lambda^m}$ belongs to
$q^{m-2l+1}A_l$.
\end{proof}

\section{The Order Bound on the Minimum Distance}
\label{sec:obound}

From now on,
$\lambda^m$ denotes the enumeration of $\Lambda^m$,
$c_m, g_m$ are the conductor and the genus of $\Lambda^m$,
$\nu^m_i$ is the $i$th value of the $\nu$-sequence corresponding
to the semigroup $\Lambda^m$
and $\delta^m_i$ is the order bound on the minimum distance of
the $i$th
one point code associated to $\mF^m$.

\begin{lemma}
\label{lemma:nu}
$\nu^1_i=i+1$ for all $i$. If $i>1$ then

{\small
$$
\nu^m_i=\left\{\begin{array}{l}\nu^{m-1}_i
\hfill 
\mbox{ if }i\leq c_m-g_m,
\\\\
\nu^{m-1}_{\frac{i+g_m}{q}-g_{m-1}}
\hfill
\begin{array}{r}
\mbox{\ \ \ \ \ \ \ \ \ \ \ \ if }c_m-g_m<i\leq 2c_m-g_m\\
\mbox{ and } q\mid i+g_m,
\end{array}
\\\\
2+2{(\lambda^m)}^{-1}(\lfloor i+g_m-c_m-1\rfloor_{\Lambda^m})
\\
\mbox{}
\hfill
\begin{array}{r}
\mbox{ if }
c_m-g_m<i\leq 2c_m-g_m\\
\mbox{ and } q\nmid i+g_m,
\end{array}
\\\\
i-g_m+1
\hfill 
\mbox{ otherwise.}
\end{array}\right.
$$
}

\end{lemma}

%
%
%

\begin{proof}
Notice that $c_m\geq qc_{m-1}$ implies that 
an element in $\Lambda^m$ is a multiple of $q$
if and only if it is in $q\Lambda^{m-1}$.

The case $i\leq c_m-g_m$ is equivalent to
$\lambda^m_i\leq c_m$.  From the inductive definition of $\Lam^m$ it
is clear that  $\lambda^m_i = q\lambda^{m-1}_i$
and consequently, $\nu_i^m=\nu_i^{m-1}$.

The case $i \geq 2c_m-g_m$ is equivalent to $\lambda^m_i \geq 2c_m$.
It is well known that for any semigroup $\Lambda$ with conductor $c$ and genus $g$
the sequence $\nu$ satisfies that $\nu_i=i-g+1$ for all $i\geq 2c-g$.

We are left with the case $c_m \leq \lambda^m_i < 2c_m$.  
Suppose $\lambda^m_i = \lambda^m_j+\lambda^m_k$ 
with $\lambda^m_j\leq \lambda^m_k$.  
Then $\lambda^m_j < c_m$ so $\lambda^m_j \in q\Lambda^{m-1}$.
If $q \mid \lambda^m_i$ then also $q \mid \lambda^m_k$, 
so $\lambda^m_k \in q\Lambda^{m-1}$.  
This shows that we have a one-to-one correspondence between 
$\{j\in\N_0:\lambda^m_i-\lambda^m_j\in\Lambda^m \}$
and 
$\{j\in\N_0:\lambda^{m-1}_{i'} -\lambda^{m-1}_j\in\Lambda^{m-1}\}$, 
where $q\lambda_{i'}^{m-1}=\lambda_i^m$, that is, $i'= (i+g_m)/q -g_{m-1}$.
Thus $\nu^m_i = \nu^{m-1}_{i'}$.

If $q \nmid \lambda^m_i$ then also $q \nmid\lambda^m_k$.
Consequently,  $\lambda^m_k > c_m$ and $\lambda^m_j <
\lambda^m_i-c_m$.
One can see that each $\lambda^m_j \in \Lambda^m$ with 
$\lambda^m_j < \lambda^m_i-c_m$ yields a pair of elements in 
$\{j\in\N_0:\lambda^m_i-\lambda^m_j\in\Lambda^m \}$.
Thus $\nu_i = 2\lvert\{\alpha \in \Lambda^m : \alpha < \lambda^m_i
-c_m\} \rvert=2+2{(\lambda^m)}^{-1}(\lfloor i+g_m-c_m-1\rfloor_{\Lambda^m})$.
  \end{proof}

\begin{lemma}
\label{lemma:obound}
The order bound on the minimum distance satisfies
{\small$$
\delta^m_i=\left\{\begin{array}{l}
2
\hfill
\mbox{ if }i\leq c_m-g_m,\\\\
\nu^m_{i+2}
\hfill
\begin{array}{r}
\mbox{\ \ \ \ \ \ \ \ \ \ \ \ if }c_m-g_m<i\leq 2c_m-g_m-2\\
\mbox{ and } q\mid i+1+g_m,
\end{array}\\\\
\nu^m_{i+1}
\hfill
\mbox{ otherwise. }
\end{array}\right.
$$}

\end{lemma}

\begin{proof}
Notice that by Lemma~\ref{lemma:nu},
$\nu^m_{c_m-g_m+1}=2$ for all $m$.
Thus,
if $i\leq c_m-g_m$
then $\delta^m_i=2$.

By Lemma~\ref{lemma:nu}, $\nu^m$ is increasing in the subset
$\{i\in{\mathbb N}_0:c_m-g_m<i\leq 2c_m-g_m, q\nmid i+g_m\}\cup\{i\in{\mathbb N}_0:i>2c_m-g_m\}.$
Now it is enough to check that for $i$ such that
$c_m-g_m< i\leq 2c_m-g_m-2$ and $q\mid i+1+g_m$,
$\nu^m_{i+1}\geq\nu^m_{i+2}$. 
To see this we will show that 
$
\{(j,k)\in{\mathbb N}_0\times{\mathbb N}_0,j\leq k:\lambda^m_j+\lambda^m_k=\lambda^m_{i+2}\}
\subseteq
\{(j,k)\in{\mathbb N}_0\times{\mathbb N}_0,j\leq k:\lambda^m_j+\lambda^m_{k-1}=\lambda^m_{i+1}\}$.
Indeed, since 
$q\mid i+1+g_m$,
$q\nmid i+2+g_m=\lambda^m_{i+2}$.
So, $\lambda^m_{i+2}=\lambda^m_j+\lambda^m_k$ with $\lambda^m_j\leq\lambda^m_k$ is only possible if
$\lambda^m_j< c_m$ and $\lambda^m_k> c_m$. In this case,
$\lambda^m_{i+1}=\lambda^m_{i+2}-1=\lambda^m_j+\lambda^m_k-1=\lambda^m_j+\lambda^m_{k-1}$.

\end{proof}

\begin{theorem}

The order bound on the minimum distance is
{\small
$$
\delta^m_i=\left\{\begin{array}{l}
2\hfill{\mbox{ if }i\leq c_m-g_m,}\\\\
\begin{array}{l}
2q^{m-\alpha}+2\frac{i+1+g_m-c_m}{q^{2\alpha-m-1}}-2c_{2m-2\alpha+1},\\
\mbox{where } \alpha=\lceil\log_q(q^m-i-1+c_m-g_m)\rceil
\end{array}
\\\hfill{\mbox{ if }
c_m-g_m< i\leq 2c_m-g_m-2,}\\\\
i-g_m+2\hfill{\mbox{\ \ \ \ \ \ \ \ \ \ \ \ \ \ \ \ \ \ \ \ if }i>2c_m-g_m-2.}
\end{array}\right.
$$
}
\\
\end{theorem}

\begin{proof}
The cases
$i\leq c_m-g_m$ and
$i>2c_m-g_m-2$
follow directly from 
Lemma~\ref{lemma:obound}
and Lemma~\ref{lemma:nu}.

Suppose $c_m-g_m<i\leq 2c_m-g_m-2$.
We will show that 
$\delta^m_i=2q^{l-1}+2\frac{i+1+g_m-c_m}{q^{m-2l+1}}-2c_{2l-1}$,
where $l=m+1-\lceil\log_q(q^m-i-1+c_m-g_m)\rceil$.
Substituting $l = m+1-\alpha$ with $\alpha$ as defined above gives the 
desired result.

If $q\mid i+1+g_m$, then
\begin{align*}
\delta^m_i & =\nu^m_{i+2} \\
&=2+2{(\lambda^m)}^{-1}(\lfloor i+1+g_m-c_m \rfloor_{\Lambda^m}) \\
&=2q^{l-1}+2\frac{i+1+g_m-c_m}{q^{m-2l+1}}-2c_{2l-1}.\\
\end{align*}
where $l=m+1-\lceil\log_q(q^m-i-1+c_m-g_m)\rceil$.
Here we have used Lemma~\ref{lemma:obound},
Lemma~\ref{lemma:nu}, and Theorem~\ref{theorem:inverseenumeration}.

For $q\nmid i+1+g_m$, 
\[ \delta^m_i=\nu^m_{i+1} =2+2{(\lambda^m)}^{-1}(\lfloor
i+g_m-c_m\rfloor_{\Lambda^m}).
\]
However, 
$\lfloor i+1+g_m-c_m\rfloor_{\Lambda^m}=\lfloor
i+g_m-c_m\rfloor_{\Lambda^m}$
since in this case  $i+1+g_m-c_m< c_m$
and $q\nmid i+1+g_m-c_m$.
Thus, we can use the same formula as above for $\del^m_i$.
\end{proof}

\def\cprime{$'$}




\end{document}